\documentclass[letter]{aa}

\usepackage{graphicx,txfonts,natbib,color}
\bibpunct{(}{)}{;}{a}{}{,}

\begin{document}

\title{Chromospheric magnetic field and density structure measurements
using hard X-rays in a flaring coronal loop}

\titlerunning{Chromospheric magnetic field and density measurements}

\author{E. P. Kontar\inst{1} \and I. G. Hannah\inst{1} \and A. L.
MacKinnon\inst{2}} \institute{Department of Physics and Astronomy,
University of Glasgow, G12 8QQ, United Kingdom\\
\email{eduard@astro.gla.ac.uk, iain@astro.gla.ac.uk}\\ \and Department
of ACE, University of Glasgow, G12 8QQ, United Kingdom\\
\email{alec@astro.gla.ac.uk}}

\date{Received July, 2008; accepted ***, 2008}

\abstract{}{A novel method of using hard X-rays as a diagnostic
for chromospheric density and magnetic structures is developed to
infer sub-arcsecond vertical variation of magnetic flux tube size and neutral
gas density.}{Using Reuven Ramaty High Energy Solar Spectroscopic
Imager (RHESSI) X-ray data and the newly developed X-ray visibilities forward
fitting technique we find the FWHM and centroid positions of hard X-ray sources with
sub-arcsecond resolution ($\sim 0.2''$) for a solar limb flare. We show
that the height variations of the chromospheric density and the magnetic
flux densities can be found with unprecedented vertical resolution of
$\sim$ 150~km by mapping 18-250~keV X-ray emission of energetic
electrons propagating in the loop at chromospheric heights of 400-1500~km.}
{Our observations suggest that the density of the neutral gas is in
good agreement with hydrostatic models  with a scale height of around
$140\pm 30$~km. FWHM sizes of the X-ray sources decrease with energy
suggesting the expansion (fanning out) of magnetic flux tube in the
chromosphere with height. The magnetic scale height
$B(z)\left(dB/dz\right)^{-1}$ is found to be of the order of 300~km and
strong horizontal magnetic field is associated with noticeable flux tube
expansion at a height of $\sim$ 900~km.}{}


\maketitle

\section{Introduction}

Chromospheric magnetic fields are notoriously difficult to measure and
their detailed structure is effectively inaccessible with modern
observations. The difficulties of various line spectroscopic techniques
\citep[e.g.][]{solanki2006} occur because the magnetic field is relatively
weak so the observed spectral lines are consequently broad and
insensitive to the field. The computation of chromospheric vector
magnetic field from spectral lines is also an ill-conditioned inverse
problem \citep[e.g.][]{metcalf1995}. In addition, current ground based
vector magnetograms have spatial resolution comparable with the vertical
size of the chromosphere itself $~2-3''$ \citep{lagg2007}. Therefore,
various indirect techniques are often employed to determine the magnetic
field in the chromosphere: optical observations of photospheric magnetic
fields combined with extrapolation into the chromosphere
\citep[e.g.][]{mcclymont1997}; radio observations of gyroresonance
emission \citep{lang1993,aschwanden1995,vourlidas1997,white1997}.

The solar chromosphere being only about 2000~km thick ($\sim 3''$)
strategically covers the layer where the solar atmosphere turns from the
gas-dominated lower chromosphere/photosphere into the magnetic field
dominated upper chromosphere/corona. \citet{gabriel1976} has proposed
that magnetic field in the chromosphere fans out (canopies) and
\citet{giovanelli1982} have found that the canopy height should be
typically 300--400 km. However, polarisation measurements by
\citet{landi1998} suggest a very small horizontal component of the magnetic
field and \citet{schrijver2003} argue that the ``wine-glass'' shaped magnetic
field should return to the photosphere near their parent flux tube. In addition,
different magnetic field models predict different canopy heights
\citep{solanki1999}.

The transport of both thermal and energetic charged particles in the solar
atmosphere is governed by individual magnetic flux tubes. Therefore flare
accelerated electrons one-dimensionally propagating along magnetic field
lines can trace individual flux tubes from the electron acceleration site in
the corona down to the deep layers of chromosphere where electrons emit
hard X-ray emission. The asymmetry of hard X-ray footpoint sizes of a
flaring loop \citep{melrose1979} could be a measure of the ratio of magnetic
field strengths in the X-ray loop footpoints \citep{schmahl2007}. The
simple dependency of X-ray emission maximum location on the photon
energy and density structure in the low corona/chromosphere due to Coulomb
collisions \citep{brown2002}, has allowed
\citet{aschwanden2002} to infer the chromospheric density structure from
Reuven Ramaty High Energy Solar Spectroscopic Imager RHESSI
\citep{lin2002} X-ray observations.

In this letter, we show that X-rays can be a diagnostic tool for the analysis of
not only energetic electrons in solar flares but of the magnetic flux tubes
and density structure in the chromosphere. We analyse the spatial
and energy distribution of hard X-ray sources using RHESSI data to infer
chromospheric density and magnetic field structure. Our results show that
the density distribution of neutral hydrogen in the flaring loop has a scale
height of 140~km and the magnetic flux tube of the flaring loop fans out by
a factor of $\sim 3$ at the height of around $900$ km.

\section{Electron precipitation in magnetic loops}

Accelerated electrons follow magnetic field lines towards the denser layers
of the atmosphere, losing energy to binary collisions and scattering in angle en route.
In the first instance we follow \citet{brown2002} and neglect scattering. Then
the flux $F(E,z)$ [electrons s$^{-1}$cm$^{-2}$keV$^{-1}$] of electrons of energy $E$
at depth $z$ from the acceleration point $z=0$, is

\begin{equation}\label{F_ez}
    F(E,z)=F_{IN}E\left(\sqrt{E^2+2KN(z)}\right)^{-\delta -1},
\end{equation}

\noindent where $N(z)$ is the column depth $N(z)=\int_0^{z}n(z)dz$,
$K=2\pi e^4 \ln\Lambda$, $e$ is the electron charge, $\ln \Lambda$ is the Coulomb
logarithm \citep{brown1973,kontar2002} and we have assumed
an injected flux of accelerated electrons $F(E,z=0)=F_{IN}E^{-\delta}$ [electrons
keV$^{-1}$ cm$^{-2}$ s$^{-1}$] at $z=0$. The hard X-ray flux from depth $z$
is

\begin{equation}\label{I_ez}
    I(\epsilon, z)=\frac{n(z)}{4\pi R^2}\int _{\epsilon}^{\infty}
    F(E,z)\sigma(\epsilon, E)dE
\end{equation}

\noindent where $\epsilon$ is the photon energy and $\sigma(\epsilon,
E)$ is isotropic bremsstrahlung cross-section \citep{haug1997}.
Since the density $n(z)$ increases with $z$ while the electron flux
$F(E,z)$ decreases, $I(\epsilon, z)$ thus exhibits a maximum for a fixed
$\epsilon$ at a depth $z_{max}(\epsilon)$ that increases monotonically
with $\epsilon$ \citep{brown2002}. Lower energy electrons lose their energy faster,
with higher energy electrons propagating deeper into the atmosphere and
the higher energy X-ray emission should come from lower layers of
the solar atmosphere.

\begin{figure} \center\includegraphics[width=60mm]{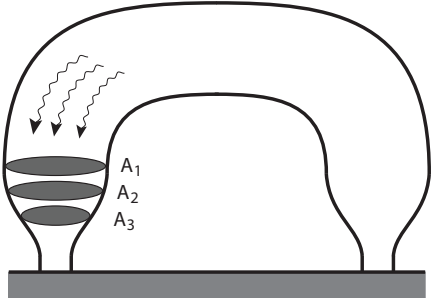}
\caption{Cartoon of a magnetic loop and electron precipitation along the
loop. The spatial size of the X-ray emission decreases with height,
indicating the changing magnetic loop width.}\label{fig:1} \end{figure}

If the magnetic field strength $B(z)$ increases with depth (Figure \ref{fig:1}),
the cross-sectional area $A(z)$ of the flux tube will obey the principle of
magnetic flux conservation $B(z)A(z)=const$
\citep[e.g.][]{melrose1979}. The size of the X-ray source at energy $\epsilon$
becomes a measure of $A(z)$ at $z_{max}(\epsilon)$ for limb events
(Figure \ref{fig:1}), which are unaffected by albedo effects \citep{kontar2006}.
We assume in the first instance that the electrons move parallel to the field
so they experience no mirror force although $B(z)$ varies.
Electrons contributing most to radiation at $z_{max}(\epsilon)$ will still be close
to their initial energies
and thus mostly  till moving along their initial directions
\citep{Brown1972,leach1981,mackinnon1991}.

\section{RHESSI X-ray observations}

We selected a large GOES M6 class X-ray flare that appeared at the
eastern limb on January 06, 2004 with a hard X-ray peak at $\sim$06:23UT.
The flare has an extended coronal source visible in soft X-rays and two hard
X-ray footpoints (Figure 2). The ``southern" footpoint is 5-10 times stronger
than the northern footpoint, which seems partially occulted at energies above
$\sim$120 keV.  This flare seems ideal for our analysis since it is a limb
event with one dominant source of hard X-ray emission seen in images up to
$\sim$300 keV. In addition, the spatially integrated count rate suggests
emission above $800$~keV. The spatially integrated photon spectrum
for the same time interval (06:22:20-06:23:00 UT) has been fitted using
isothermal plus thick-target model (Figure 3). The X-ray emission above
18 keV (Figure 3) is dominated by the footpoint thick-target emission and
can be used for our analysis. At these energies the ``southern'' X-ray
source is the brightest and is used for detailed imaging analysis in different
energy ranges.

\begin{figure}
\center\includegraphics[width=66mm]{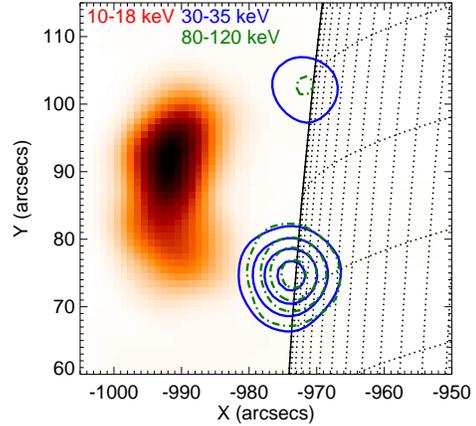}
\caption{RHESSI CLEAN \citep{hurford2002} X-ray images of the January 6,
2004 limb event. The contours show hard X-ray emission integrated
for the impulsive phase of the flare (06:22:20-06:23:00 UT) from the
footpoints in 30-35 keV (solid blue line) and 80-120 keV (dot-dashed green
lines). The background image shows subsequent softer thermal emission
(06:24:00-06:24:40 UT) in 10-18 keV.}\label{fig:2} \end{figure}

\begin{figure}
\center\includegraphics[width=66mm]{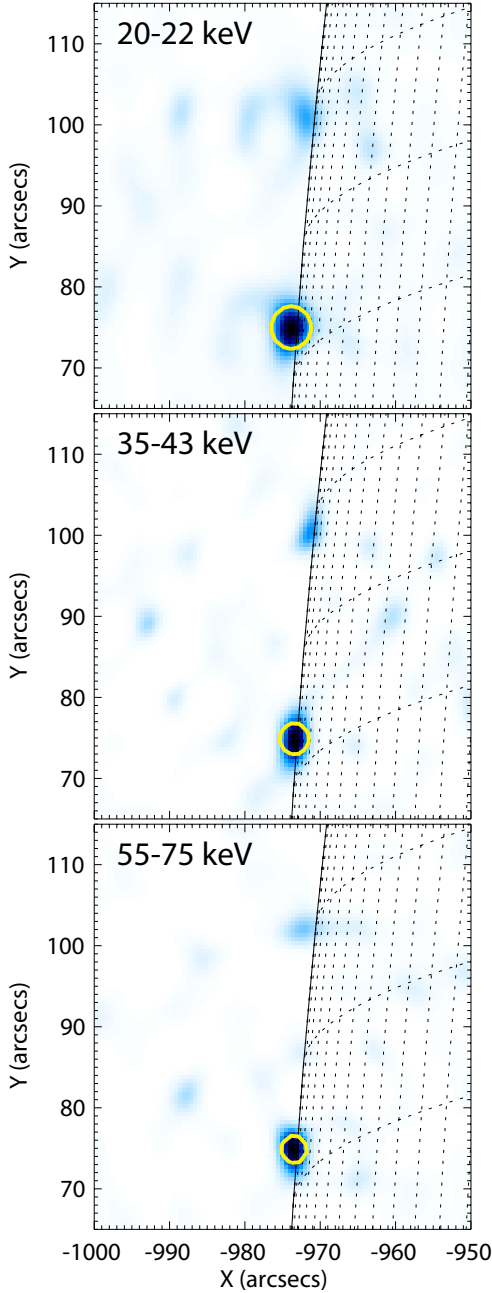}
\caption{RHESSI MEM NJIT \citep{schmahl2007} images in 3 energy ranges
(top to bottom: 20-22 keV, 35-43 keV, 55-75 keV) integrated for the
impulsive phase of the flare (06:22:20-06:23:00 UT). The FWHM circular
source size found by forward fitting the X-ray visibilities, which were also
used to make the background images, are shown by the overplotted
contours (orange lines).}\label{fig:2} \end{figure}

X-ray visibilities (2-dimensional spatial Fourier components)
\citep[see][]{schmahl2007} in ten different hard X-ray energy ranges from
18 to 250 keV have been forward fitted using single circular Gaussian
source (Figure 3). Although the figure shows the fitted source matching the
imaged footpoint location, the fit has been conducted on the X-ray
visibilities from which the images are derived, not the images themselves.
The X-ray images \emph{per se} are generally a poor indicator of the
source size \citep{hurford2002,emslie2003}. Visibility forward fit however
provides reliable spatial measures and clear statistical uncertainties for
all the fit parameters. These fits provided us with the centroid positions of
the Gaussian source ($x(\epsilon)$, $y(\epsilon)$), and its Full Width
Half-Maximum (FWHM) size $s(\epsilon)$. Assuming vertically emerging
magnetic field and calculating the radial
distance measured from the disk centre of the Sun
$R(\epsilon)=\sqrt{x(\epsilon)^2+y(\epsilon)^2}$ we can readily find the
height of the X-ray source $h(\epsilon)=R(\epsilon)-R_{0}$, where $R_0$ is
radial distance of the bottom of the loop.  The typical uncertainties on
radial distance measurement $R(\epsilon)$ are around $0.2''$ $\sim150$ km. $R_0$ is poorly
known \citep{aschwanden2002} but crucial for the
detailed analysis, therefore we incorporate it as a fit parameter.
Assuming the neutral hydrogen density profile to be
\begin{equation}\label{n_h}
    n(h(\epsilon))=n_0\exp\left(-(R(\epsilon)-R_0)/h_0\right)
    \label{n_h}
\end{equation}
\noindent where $h_0$ is the density scale height, and $n_0$ is the number
density at height $h =0$. Various chromospheric models A-F from
\citet{vernazza1981} provide slightly different values of $n_0$ though most are
very close to $n_0=1.16\times 10 ^{17}$~cm$^{-3}$. Therefore we use this $n_0$
as a boundary condition to find the characteristic scale-height as well as
$R_0$. Using equation (\ref{n_h}) to fit the maximum of flux spectrum given by
Equation (\ref{I_ez}) with $\ln \Lambda=\ln \Lambda _{eH}= 7.1$
\citep{brown1973,kontar2002} and using the spectral index found from the
spatially integrated spectrum $\delta = 3.2$ (top panel Figure 4) we find
$R_0=975.3\pm 0.2''$ and $h_0=140\pm30$~km. Using these fitted parameters we can
now plot the centroid height $h$ against energy (Figure 4, middle panel), with
the fitted model overplotted, finding that the height decreases by less
than $500$~km between $20$~keV to $\sim 200$~keV.

The FWHM sizes $s(\epsilon)$ of the source, found from forward fitting
the visibilities, decrease with energy from $6.2''$ (4.5~Mm) down to $2.3''$
(1.7~Mm) (Figure 4, bottom panel). Estimation of the magnetic field
structure and density structure is shown in the top and bottom panels of
Figure 5 respectively. We find that the magnetic field structure widens
with increasing height and that the density decreases by over 2 orders of
magnitude from a height of 1~Mm above the photosphere.

\begin{figure} \center \includegraphics[width=68mm]{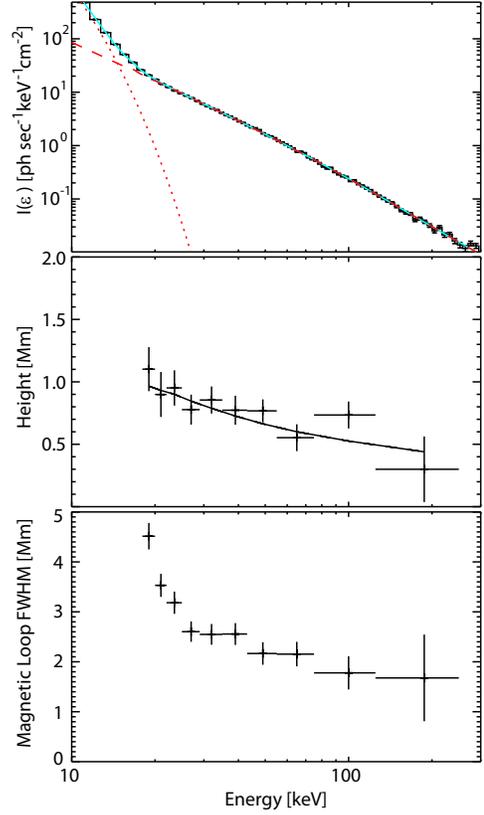}
\caption{Top: Spatially integrated RHESSI X-ray spectrum (histogram) of
the flare's impulsive phase (06:22:20-06:23:00 UT) fitted with an
iso-thermal plus thick target model fit (solid line). The fit provides a
temperature $1.65$~keV (19MK), emission measure $2\times 10^{48}$
cm$^{-3}$(dotted lines) and spectral indices $\delta=3.2$ below 191~keV
and 3.9 above. Middle:``Southern" hard X-ray source height as a function
of energy, $h(\epsilon)=R(\epsilon)-R_{0}$, where $R_{0}=975.3''$;
Bottom: FWHM circular source size as a function of energy.}\label{fig:4}
\end{figure}

 \begin{figure}
 \center\includegraphics[width=66mm]{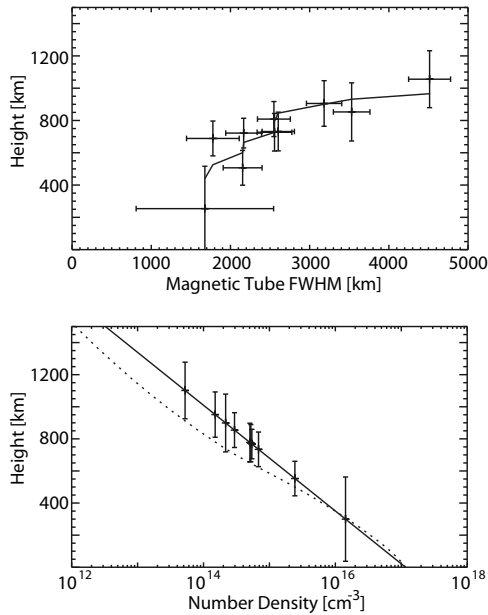}
\caption{Top: Chromospheric magnetic flux tube FWHM at various heights with
FWHM for density fit given by Equation \ref{n_h}.
(solid line) Bottom: The chromospheric neutral number
hydrogen density as a function of height; best fit
with hydrogen density profile given by Equation (\ref{n_h}) (solid line),
density model C from \citet{vernazza1981} (dashed line).}\label{fig:5}
\end{figure}

\section{Discussion and Conclusions}

Forward fitting X-ray visibilities allow simple and reliable measurements of
not only locations of the emission maxima but also the characteristic sizes
of hard X-ray sources. By using this visibility analysis on a good candidate
limb flare we have improved earlier chromospheric height and density
measurements of \citet{aschwanden2002}, reducing the uncertainty of
emission maximum positions down to $\sim0.2''$. Assuming collisional transport in neutral
hydrogen it can be concluded that the chromospheric density is consistent
with a gravitationally stratified atmosphere of density scale height
$140\pm 30$~km. We show for the first time that not only is the higher
energy X-ray emission produced continuously deeper in the
chromosphere but the X-ray source sizes decrease from $\sim6.2''$
down to $\sim2.3''$. The precise measurement of the characteristic sizes allows
us to conclude  that magnetic field directing the energetic electrons
converge, with the magnetic flux tube shrinking from
FWHM $\sim3.5$~Mm at $h\sim 1$~Mm down to FWHM $\sim2.5$~Mm at $h\sim 0.8$~Mm above
the photosphere. The magnetic scale height estimated using flux conservation $B(h)\sim FWHM(h)^{-2}$
as the ratio of areas $B(h)\left(dB/dh\right)^{-1}= -FWHM(h)\mbox{d}h/\mbox{d}FWHM(h)/2$
is found to be of the order of 300~km. This confirms the fanning out (canopies) of the chromospheric
magnetic fields \citep{gabriel1976}.

Our deduced density structure agrees quite well with independent
estimations from other methods, possibly a surprising conclusion in view of
the simple treatment of electron transport, completely neglecting any
pitch-angle changes. As already noted, collisional scattering will modify
the range only by a factor of order unity but magnetic moment
conservation might have a greater effect, unless fast electrons all start
with velocity vectors parallel to $\vec{B}$. Large pitch angles would develop
via collisional scattering only as electrons reached the end of their ranges,
and the mirror force would be similarly unimportant most of the time. Such
a concentration at small pitch angles seems at odds with findings of a
nearly isotropic electron distribution from studies of photospheric albedo
\citep{KontarBrown06mirror}. The magnetic field convergence we find here however
offers a simple resolution. The flux tube implied by our HXR
source FWHM's implies magnetic field lines substantially inclined to the
vertical, possibly by as much as 60\% on average at the outer edge and in
addition are likely to be twisted. The HXR polar diagram of electrons
populating the whole of an ensemble of field lines, with such a range of
angles to the vertical, would be much closer to isotropic than expected for
vertical field lines. The presence of magnetic canopies, then, appears to
be a critical factor for interpreting HXR images and even spectra, but one
that needs a much more substantial and detailed modelling effort than we
attempt here.

While the X-ray diagnostics of the chromospheric magnetic field is an
extremely attractive new technique, we emphasise that the morphology of
X-ray sources should be further scrutinised. We note that the shape of the
hard X-ray sources is more elliptical (Figure 3) rather than circular
suggesting that the vertical extend of X-ray sources is govern by the
spatially varying magnetic field (convergence and a twist of the flux tube),
through the conservation of the electron magnetic moment. This
complicates the interpretation of the HXR source sizes beyond the simple
ideas embodied in (\ref{F_ez}) and (\ref{I_ez}) but brings a new
diagnostic potential. This requires a more complicated electron dynamic
and forward fitting elliptical sources to the X-ray visibilities, which will be
the subject of future work.

The method of measuring magnetic field using X-rays could be viewed as unique
tool of loop width measurements in the chromosphere comparable to using TRACE
data to measure the widths of flux tubes in the corona \citep{watko2000}. While
it is often argued that the heights of line formation should not be assigned due
to the fact that very distinct layers of the atmosphere could be sampled
\citep[e.g.][]{sanchez1996}, X-rays are uniquely related to the magnetic field
lines connecting the photosphere and the electron injection site in the corona
and weakly sensitive (via Coulomb logarithm dependency) to the temperature
variations in the chromosphere. Therefore hard X-ray emission is likely to be
a valuable diagnostic for mapping chromospheric magnetic field and density
structures.

\begin{acknowledgements}

This work was supported by a STFC rolling grant (EPK, IGH, ALM) and
STFC/PPARC Advanced Fellowship (EPK). Financial support
by the European Commission through the SOLAIRE Network
(MTRN-CT-2006-035484) is gratefully acknowledged. 

\end{acknowledgements}

\bibliographystyle{aa} \bibliography{refs}

\end{document}